\documentstyle[epsfig]{aprim}

\newif\ifAMStwofonts

\ifoldfss
  \ifCUPmtlplainloaded \else
    \NewTextAlphabet{textbfit} {cmbxti10} {}
    \NewTextAlphabet{textbfss} {cmssbx10} {}
    \NewMathAlphabet{mathbfit} {cmbxti10} {} 
    \NewMathAlphabet{mathbfss} {cmssbx10} {} 
  \fi
  \ifAMStwofonts
    \ifCUPmtlplainloaded \else
      \NewSymbolFont{upmath} {eurm10}
      \NewSymbolFont{AMSa} {msam10}
      \NewMathSymbol{\upi}     {0}{upmath}{19}
      \NewMathSymbol{\umu}     {0}{upmath}{16}
      \NewMathSymbol{\upartial}{0}{upmath}{40}
      \NewMathSymbol{\leqslant}{3}{AMSa}{36}
      \NewMathSymbol{\geqslant}{3}{AMSa}{3E}

    \fi
  \fi
\fi 

\ifnfssone
  \newmathalphabet{\mathit}
  \addtoversion{normal}{\mathit}{cmr}{m}{it}
  \addtoversion{bold}{\mathit}{cmr}{bx}{it}
  \newmathalphabet{\mathbfit} 
  \addtoversion{normal}{\mathbfit}{cmr}{bx}{it}
  \addtoversion{bold}{\mathbfit}{cmr}{bx}{it}
  \newmathalphabet{\mathbfss} 
  \addtoversion{normal}{\mathbfss}{cmss}{bx}{n}
  \addtoversion{bold}{\mathbfss}{cmss}{bx}{n}
  \ifAMStwofonts
    \ifCUPmtlplainloaded \else
      \UseAMStwoboldmath
      \makeatletter
      \new@mathgroup\upmath@group
      \define@mathgroup\mv@normal\upmath@group{eur}{m}{n}
      \define@mathgroup\mv@bold\upmath@group{eur}{b}{n}
      \edef\UPM{\hexnumber\upmath@group}
      \new@mathgroup\amsa@group
      \define@mathgroup\mv@normal\amsa@group{msa}{m}{n}
      \define@mathgroup\mv@bold\amsa@group{msa}{m}{n}
      \edef\AMSa{\hexnumber\amsa@group}
      \makeatother
      \mathchardef\upi="0\UPM19
      \mathchardef\umu="0\UPM16
      \mathchardef\upartial="0\UPM40
      \mathchardef\leqslant="3\AMSa36
      \mathchardef\geqslant="3\AMSa3E
    \fi
  \fi
\fi 

\ifnfsstwo
  \DeclareMathAlphabet{\mathbfit}{OT1}{cmr}{bx}{it}
  \SetMathAlphabet\mathbfit{bold}{OT1}{cmr}{bx}{it}
  \DeclareMathAlphabet{\mathbfss}{OT1}{cmss}{bx}{n}
  \SetMathAlphabet\mathbfss{bold}{OT1}{cmss}{bx}{n}
  \ifAMStwofonts
    \ifCUPmtlplainloaded \else
      \DeclareSymbolFont{UPM}{U}{eur}{m}{n}
      \SetSymbolFont{UPM}{bold}{U}{eur}{b}{n}
      \DeclareSymbolFont{AMSa}{U}{msa}{m}{n}
      \DeclareMathSymbol{\upi}{0}{UPM}{"19}
      \DeclareMathSymbol{\umu}{0}{UPM}{"16}
      \DeclareMathSymbol{\upartial}{0}{UPM}{"40}
      \DeclareMathSymbol{\leqslant}{3}{AMSa}{"36}
      \DeclareMathSymbol{\geqslant}{3}{AMSa}{"3E}
    \fi
  \fi
\fi 

\ifCUPmtlplainloaded \else
  \ifAMStwofonts \else 
    \def\upi{\pi}
    \def\umu{\mu}
    \def\upartial{\partial}
  \fi
\fi

\title[Resonant Kuiper Belt Objects]
 {On the Dynamics of Resonant Kuiper Belt Objects}

\author[Jiang and Yeh]
       {Ing-Guey Jiang$^1$ and Li-Chin Yeh$^2$\\
  $^1$Institute of Astronomy, National Central University, Chung-Li, Taiwan\\
  $^2$Department of Applied Mathematics, National Hsinchu University of Education, Hsin-Chu, Taiwan}
\date{}

\pagerange{\pageref{firstpage}--\pageref{lastpage}}
\pubyear{2005}

\begin{document}

\maketitle

\label{firstpage}

\begin{abstract}

We propose a new mechanism of drag-induced resonant capture, which can
explain the resonant Kuiper Belt Objects 
in a natural way. A review and comparison 
with the traditional mechanism of sweeping capture by the migrating Neptune
will be given.

\end{abstract}


\section{Introduction}

The Kuiper Belt is the small bodies that in habit the outer Solar System
beyond the current orbit of Neptune.
The discovery of Kuiper Belt Objects (Jewitt and Luu 1993) has refreshed
the richness of the Solar System study. 
Particularly, the dynamical evolution of Kuiper Belt Objects (KBOs) 
probably preserves
the important information of the Solar System's early history.

There are many interesting problems about the Kuiper Belt. 
For example, Allen, Bernstein and Malhotra (2001) claimed it is likely
that there is a sharp edge around 50 AU for the Kuiper Belt. 
How this sharp edge could form is a big puzzle. 
Melita, Larwood and Williams (2005) considered a close stellar fly-by 
as an explanation for the abrupt termination at around 50 AU.
Kenyon and Bromley (2004) used the similar idea to explain the Sedna's orbit.

On the other hand, the total mass of the Kuiper Belt is a controversial issue. 
In order to form 100 km size KBOs around the current region, one would 
need a much more massive Kuiper Belt initially
because the planetesimal accretion rate has to be high enough
to make KBOs formed in time, particularly for those even larger KBOs, 
say Sedna. 
However, the material in the Kuiper Belt has to be depleted 
significantly during the evolution 
because the observational upper limit of the current mass at the Kuiper 
Belt is only 0.1 Earth mass.  

The KBOs are grouped into several classes according to 
their dynamical properties.
The way of grouping might keep changing because the number 
of detected KBOs increases quickly.

Conventionally, there are three classes of KBOs. One of the population are the
Resonant KBOs. They are trapped in 3:2 mean-motion resonance with Neptune
at 39.4 AU. About one-third of the total population are in this class.
There are also KBOs trapped into other resonance but much less populated than
3:2 resonance.
The second class mainly occupies the region between 41 AU and 46 AU with small
eccentricities. They are called the Classical KBOs.
There are also  Scattered KBOs and they are moving on
the large semi-major-axis, highly eccentric, and inclined orbits.
These KBOs could originate from the scattering
by Neptune (Luu et al. 1997).

\section{The Orbital Stability}

In general, 
the most complete numerical survey on the orbital stability in the Kuiper Belt
is done in Duncan et al. (1995).   
Their long-term intergrations cover a wide range in orbital element space
and therefore provide a good reference for the orbital stability and 
evolution around the Kuiper Belt region.

In addition to that, 
there are also many other dynamical studies 
focusing on different topics about KBOs.
\section{The Resonant Kuiper Belt Objects}

It is particularly interesting that one-third of KBOs are in 3:2 but not 
that 
many in other resonances. We discuss the possible mechanisms below: both the
conventional one and the newly proposed one in Jiang \& Yeh (2004).
 
\subsection{The Neptune's Migrations}
It was suggested by Fernandez \& Ip (1984) that the orbits of the giant 
planets may have shifted after the solar nebula had dissipated. 
Malhotra (1995) assumed that the Neptune migrates outward following a 
formula in which the semimajor axis increases with time and the increasing 
speed decays exponentially while the orbit remains to be circular.
As long as the total expansion of semimajor axis is at least 5 AU,
this model can capture KBOs into 3:2 resonance.

As described in the recent work by Hahn \& Malhotra (2005), 
in addition to the strengths
there are weaknesses for this picture.
The strength is that it can capture KBOs into 3:2 resonance 
by considering the gravity only. It is a pure dynamical effect,
so the model is clear and simple. It might also explain the Pluto's
inclination etc. 

Let us discuss the weaknesses here. 
One possible question would be about the way of Neptune's migration.
Could Neptune really migrate like that ? Is that a good approximation ? 
In the numerical simulations in Thommes et al. (1999),
the orbital eccentricity increases as the Neptune migrates, which is consistent
with the analytic results in Yeh \& Jiang (2001).
To become nearly circular in the end, 
as shown in Thommes et al. (1999), one would need a massive
Kuiper Belt to interact with the Neptune. It is unclear how massive
the Kuiper Belt could be during the epoch of early Solar System.

Another problem is about the population of 2:1 resonance.
As shown in Hahn \& Malhotra (2005), this model predicts a 2:1 resonance 
that is 2.5 times more abundant than the 3:2, while the observations show
a 2:1 resonance which is sparsely populated. 
Zhou et al. (2002) solved this problem by introducing a stochastic term
in the function which describes the increasing of semimajor axis. 
However, how could the orbit
still remain to be circular 
when the Neptune is kicked randomly would be an interesting question.

\subsection{The Drag-Induced Resonant Capture}

Jiang \& Yeh (2004) used a model of the disc-star-planet system
to investigate the effect of a gaseous proto-stellar disc on the resonance 
capture. In addition to the force from the central star and planet, the test 
particle is also influenced by the gravitational and drag forces
from the disc.
The model is used to study the problem of resonant KBOs and focus
on the 3:2 and 2:1 resonances.

Their results show that the drag force plays an important role for 
the resonant capture and many KBOs can get captured into the 
3:2 resonance but not into the
2:1 resonance. 
This is consistent with the observation that about 
one-third of the whole population
of KBOs are in the 
3:2 resonance and only a few are claimed to be in the 2:1 resonance.

The speed of inward migration shall be proportional to 
the strength of the drag force.
Thus, the stronger force could make more particles migrate and get captured
by the 3:2 resonance given that 
there are enough particles in the outer part of the Kuiper Belt. 
Therefore, the results could be related to the  
particles' initial distribution and also the strength of the drag force.

Because the proto-stellar disc definitely exists and provides the drag force
when the proto-KBOs are forming, the mechanism of drag-induced resonant
capture seems to be attractive and cannot be ignored.
It explains the resonant KBOs in a natural way.

The major weakness of this mechanism is that whether the drag or viscosity
is really as large as the one assumed in Jiang \& Yeh (2004). 
The formula they used for the drag force is equivalent to the Epstein drag
(Youdin \& Chiang 2004) given that the particles' sizes are assumed to be 
similar. The values of sound speed and size etc. in the 
Epstein drag are absorbed into one parameter $\alpha$ in Jiang \& Yeh (2004).
According to Youdin \& Chiang (2004), the stopping timescale of this drag is 
short so the drag is very important. 

It is unclear that what the sound speed exactly 
is at the outer Solar System during
the resonant capture. Nevertheless, there might be a time that the sound
speed is high enough to provide gaseous drag for these debris dusty 
particles
while they are growing to be proto-KBOs. This drag would then trigger 
the process demonstrated in Jiang \& Yeh (2004). These debris particles 
or proto-KBOs would then keep growing to be the current size KBOs even after
they are already captured into the resonance.

\section{Concluding Remarks}

The dynamical evolution of KBOs remains to be a 
very interesting problem, which is complicated by the detail of KBOs' 
formation. In particularly, the resonant KBOs will need much more 
investigations, both observationally and theoretically.

The real picture could be that, the Neptune migrate slightly during its
formation, though less than 5 AU, it might still capture some proto-KBOs into 
3:2 resonance. Simultaneously, the drag acts as another important mechanism to 
make more proto-KBOs captured into 3:2. In this way, 
there would be only a few to be in 2:1 resonance.

The detail of the mechanism of drag-induced resonant capture
and its combination with the traditional sweeping capture by the 
migrating Neptune
would be an interesting work.

\section*{Acknowledgment}
We are grateful to the National Center for High-performance Computing
for computer time and facilities.

\label{lastpage}

\clearpage

\end{document}